\documentclass[12pt,letterpaper,american,english,aps]{revtex4-2}
\usepackage[T1]{fontenc}
\usepackage[latin9]{inputenc}
\usepackage{amsmath}
\usepackage{amssymb}
\usepackage{graphicx}

\makeatletter

\pdfpageheight\paperheight
\pdfpagewidth\paperwidth

\usepackage{babel}

\makeatother

\usepackage{babel}
\begin{document}
\title{Fermionic model of unitary transport of qubits from a black hole}
\author{Bogus\l aw Broda}
\email{boguslaw.broda@uni.lodz.pl}

\affiliation{\selectlanguage{american}%
Department of Theoretical Physics, Faculty of Physics and Applied
Informatics,University of \L\'od\'{z}, 90-236 \L\'od\'{z}, Pomorska
149/153, Poland}
\homepage{http://merlin.phys.uni.lodz.pl/BBroda}

\begin{abstract}
Inspired by a recent model of Osuga and Page, we propose an explicitly
unitary fermionic toy model for transferring information from a black
hole to the outgoing radiation. The model treats the unitary evolution
as a composition of the Hawking pair creation outside the black hole
and of pair annihilation inside the black hole.

\newpage{}
\end{abstract}
\maketitle

\section{Introduction}

The black hole (BH) information (loss) paradox concerns difficulties
around the issue of unitarity of BH evaporation (for recent reviews
see e.g.\ \citep{Polchinski2017c,Harlow2016a,Chakraborty2017e,Marolf2017a}).
There are a lot of approaches proposed to date to analyze and resolve
the paradox --- some of them suggest to study simplified situations
embodied in various qubit models (see e.g.\ \citep{Giddings2012a,Avery2013,Osuga2018,Giddings2013d,Giddings2012,Mathur2009,Mathur2009b,Mathur2011}).
A successful model of BH evolution should include description of particle
pair production according to the Hawking prescription, following gradual
evaporation (``vanishing'') of the BH, and it should be unitary.
A model strictly motivated by actual physical phenomena would certainly
be greatly appreciated, but in fact any model respecting at least
general physical laws, even without any real physical mechanism built
in, would be welcome as a ``proof of concept''.

Recently Osuga and Page \citep{Osuga2018} (inspired by \citep{Almheiri2013})
have proposed an explicitly unitary toy qubit transport model for
BH evaporation (without firewalls). Another version of the model (with
additional features) has been presented in \citep{Broda2020}. In
the present paper, building on the both models, we propose yet another
toy qubit transport model for BH evaporation, which is explicitly
unitary. Since the model, by assumption, operates on qubits and particle
pair production scheme exactly follows the Hawking mechanism for fermions,
we shall work in terms of fermionic modes rather than bosonic ones.
A new and important feature of our present proposal is explicit incorporation
of the (fermionic) Hawking pair creation mechanism into the chain
of unitary processes. In other words, the global unitary evolution
considered is given by the composition $U=U''\cdot U'$, where $U'$
corresponds to creation of fermionic pairs outside a BH according
to the Hawking prescription, whereas $U''$ corresponds to annihilation
of fermion pairs inside the BH (as described in \citep{Broda2020}).

For reader's convenience, we will follow notation of \citep{Osuga2018}
(and \citep{Broda2020}) as closely as possible.

\section{The toy model}

An initial total quantum state describing a newly formed (fermionic)
BH and ``fermionic radiation'' in the vacuum state is assumed in
the following (partially product) form \citep{Broda2020} (cf. \citep{Osuga2018})
\begin{equation}
\left|\Psi\right>=\sum_{q_{1},q_{2},\dots,q_{N}=0}^{1}A_{q_{1}q_{2}\cdots q_{N}}\bigotimes_{k=1}^{N}\left|q_{k}\right>_{a_{k}}\otimes\left|\text{O}\right>_{b_{k}c_{k}}.\label{eq:initial_state}
\end{equation}
Here $A_{q_{1}q_{2}\cdots q_{N}}$ are amplitudes for inner BH modes
$a_{k}$, which encode a quantum state of the BH, and the vacuum state
for ``fermionic radiation'' is 
\begin{equation}
\left|\text{O}\right>_{b_{k}c_{k}}\equiv\left|0\right>_{b_{k}}\otimes\left|0\right>_{c_{k}},\label{eq:vacuum_state}
\end{equation}
where the Hawking (fermionic) modes $b_{k}$ and $c_{k}$ are infalling
and outgoing modes, respectively. The same range of indices ($k=1,2,\dots,N$)
postulated for BH modes $a_{k}$ and $b_{k}$, $c_{k}$ pairs is not
only a convenient computational simplification in our model but also
it is a physically justified assumption, at least approximately (see
e.g.\  \citep{Broda2020}).

In the language of $k$-mode blocks, the first step of (unitary) evolution
denoted by $U'_{k}$ yields the Hawking (fermion) pair for a single
``$k$'' mode, i.e. 
\begin{equation}
U'_{k}\left(\left|q_{k}\right>_{a_{k}}\otimes\left|\text{O}\right>_{b_{k}c_{k}}\right)=\left|q_{k}\right>_{a_{k}}\otimes\left|\text{H}_{1}\right>_{b_{k}c_{k}},\label{eq:imode_primeu}
\end{equation}
where the fermionic Hawking state can be chosen in the form (cf. Eq.(116)
in \citep{Mann:2015luq}) 
\begin{equation}
\left|\text{H}_{1}\right>_{b_{k}c_{k}}\equiv\cos\omega_{k}\left|0\right>_{b_{k}}\otimes\left|0\right>_{c_{k}}+\sin\omega_{k}\left|1\right>_{b_{k}}\otimes\left|1\right>_{c_{k}},\label{eq:hawking_state}
\end{equation}
with $\omega_{k}$ determined by BH parameter(s).

The total (i.e. for all modes $k$) $U'$-evolution yields by virtue
of ({\ref{eq:imode_primeu}}) the total (intermediate) state 
\begin{equation}
\left|\Psi'\right>=\sum_{q_{1},q_{2},\dots,q_{N}=0}^{1}A_{q_{1}q_{2}\cdots q_{N}}\bigotimes_{i=1}^{N}\left|q_{k}\right>_{a_{k}}\otimes\left|\text{H}_{1}\right>_{b_{k}c_{k}}.\label{eq:intermediate_state}
\end{equation}
We could possibly consider a slight generalization of the unitary
evolution ({\ref{eq:imode_primeu}}) allowing some unitary transformation
$\mathcal{U}'_{k}\left|q_{k}\right>_{a_{k}}$ of the internal BH mode
$a_{k}$ on the RHS of ({\ref{eq:imode_primeu}}), but we ignore
this option, because it would merely give rise to a redefinition of
$A$-amplitudes ($A_{q_{1}q_{2}\cdots q_{N}}\longmapsto A_{q_{1}q_{2}\cdots q_{N}}'$)
in the final state.

The second step of (unitary) evolution denoted by $U''_{k}$ yields
particle pair annihilation inside the BH (see \citep{Broda2020} and
cf. Eq. (3.3) in \citep{Osuga2018,Almheiri2013}), i.e. 
\begin{equation}
U''_{k}\left(\left|q_{k}\right>_{a_{k}}\otimes\left|\text{H}_{1}\right>_{b_{k}c_{k}}\right)=\left|\text{O}\right>_{a_{k}b_{k}}\otimes\left|q_{k}\right>_{c_{k}},\label{eq:imode_bisu}
\end{equation}
where the vacuum state $\left|\text{O}\right>_{a_{k}b_{k}}$ is defined
analogously to ({\ref{eq:vacuum_state}}) (with appropriate replacements
of modes).

Consequently, for the entire evolution $U_{k}\equiv U_{k}''\cdot U_{k}'$
we have 
\begin{equation}
U_{k}\left(\left|q_{k}\right>_{a_{k}}\otimes\left|\text{O}\right>_{b_{k}c_{k}}\right)=\left|\text{O}\right>_{a_{k}b_{k}}\otimes\left|q_{k}\right>_{c_{k}},\label{eq:imode_u}
\end{equation}
and the final total state assumes the form 
\begin{equation}
\left|\Psi''\right>=\sum_{q_{1},q_{2},\dots,q_{N}=0}^{1}A_{q_{1}q_{2}\cdots q_{N}}\bigotimes_{k=1}^{N}\left|\text{O}\right>_{a_{k}b_{k}}\otimes\left|q_{k}\right>_{c_{k}}.\label{eq:final_state}
\end{equation}
Eq. ({\ref{eq:final_state}}) means that all information has been
transferred from a BH to the outgoing radiation, and the BH is in
the vacuum state.

Obviously, the total operators $U'$, $U''$, $U$ are the following
tensor products of the above-defined $k$-mode operators 
\begin{equation}
U'=\bigotimes_{k=1}^{N}U'_{k},\quad U''=\bigotimes_{k=1}^{N}U''_{k},\quad U=\bigotimes_{k=1}^{N}U''_{k}\cdot U'_{k}\equiv\bigotimes_{k=1}^{N}U_{k},\label{eq:operatorsu_as_products}
\end{equation}
respectively, and 
\begin{equation}
U'\left|\Psi\right>=\left|\Psi'\right>,\quad U''\left|\Psi'\right>=\left|\Psi''\right>,\quad U\left|\Psi\right>\equiv U''\cdot U'\left|\Psi\right>=\left|\Psi''\right>.\label{eq:opearatorsu_onstates}
\end{equation}

\section{Unitary operators}

We shall now explicitly derive the implicitly defined unitary operators
$U'_{k}$, $U''_{k}$ and $U_{k}$. To this end let us first observe
the following elementary fact from linear algebra: namely, for each
pair of orthonormal bases $\left\{ \left|E_{\Lambda}\right>\right\} $,
$\left\{ \left|E_{\Lambda}'\right>\right\} $ ($\left<E_{\Lambda}|E_{\Lambda'}\right>=\left<E_{\Lambda}'|E_{\Lambda'}'\right>=\delta_{\Lambda\Lambda'}$)
in a finite dimensional Hilbert space $\mathcal{H}$ we can construct
an operator 
\begin{equation}
U=\sum_{\Lambda}\left|E_{\Lambda}'\right>\left<E_{\Lambda}\right|,\label{eq:general_u}
\end{equation}
which is explicitly unitary. Really, we can easily check that e.g.
\begin{equation}
U^{\dagger}\cdot U=\sum_{\Lambda,\Lambda'}\left|E_{\Lambda}\right>\left<E_{\Lambda}'|E_{\Lambda'}'\right>\left<E_{\Lambda'}\right|=\sum_{\Lambda,\Lambda'}\delta_{\Lambda\Lambda'}\left|E_{\Lambda}\right>\left<E_{\Lambda'}\right|=\mathbb{I}.\label{eq:unitarity_of_u}
\end{equation}
Since the total Hilbert space $\mathcal{H}$ is a tensor product of
$N$ $k$-mode Hilbert spaces $\mathcal{H}_{k}$, i.e. $\mathcal{H}=\otimes_{k=1}^{N}\mathcal{H}_{k}$,
we can confine our construction to the single $k$-mode space $\mathcal{H}_{k}=\mathcal{H}_{a_{k}}\otimes\mathcal{H}_{b_{k}}\otimes\mathcal{H}_{c_{k}}$,
where $\dim_{\mathbb{C}}\mathcal{H}_{k}=2\cdot2\cdot2=8$. Then, our
unitary operators will be defined by three 8-dimensional orthonormal
($k$-dependent) bases in $\mathcal{H}_{k}$.

The first base, $\left\{ \left|E_{\Lambda}\right>_{k}\right\} _{\Lambda=0}^{7}$,
assumes the following standard form: 
\begin{equation}
\begin{array}{rcl}
\left|E_{0}\right>_{k} & = & \left|0\right>_{a_{k}}\otimes\left|0\right>_{b_{k}}\otimes\left|0\right>_{c_{k}}\\
\left|E_{1}\right>_{k} & = & \left|0\right>_{a_{k}}\otimes\left|0\right>_{b_{k}}\otimes\left|1\right>_{c_{k}}\\
 & \vdots\\
\left|E_{7}\right>_{k} & = & \left|1\right>_{a_{k}}\otimes\left|1\right>_{b_{k}}\otimes\left|1\right>_{c_{k}}
\end{array}\label{eq:base_e_defined}
\end{equation}
(or $\left|E_{\Lambda}\right>_{k}=\left|\Lambda\right>_{a_{k}b_{k}c_{k}}$,
in short).

The second ($\tau$-dependent) base, $\left\{ \left|E_{\Lambda}'\left(\tau\right)\right>_{k}\right\} _{\Lambda=0}^{7}$,
is given by (for further convenience, we have also included expansions
in terms of $\tau$): 
\begin{equation}
\begin{array}{rcl}
\left|E_{0}'\left(\tau\right)\right>_{k} & = & \left|0\right>_{a_{k}}\otimes\left[\cos\left(\omega_{k}\tau\right)\left|0\right>_{b_{k}}\otimes\left|0\right>_{c_{k}}+\sin\left(\omega_{k}\tau\right)\left|1\right>_{b_{k}}\otimes\left|1\right>_{c_{k}}\right]\\
 &  & \equiv\left|0\right>_{a_{k}}\otimes\left|\text{H}_{\tau}\right>_{b_{k}c_{k}}\\
 &  & \equiv\cos\left(\omega_{k}\tau\right)\left|E_{0}\right>_{k}+\sin\left(\omega_{k}\tau\right)\left|E_{3}\right>_{k}\\
 &  & \quad=\left|E_{0}\right>_{k}+\omega_{k}\tau\left|E_{3}\right>_{k}+\mathcal{O}(\tau^{2})\\
\left|E_{1}'\right>_{k} & = & \left|0\right>_{a_{k}}\otimes\left|0\right>_{b_{k}}\otimes\left|1\right>_{c_{k}}\equiv\left|E_{1}\right>_{k}\\
\left|E_{2}'\right>_{k} & = & \left|0\right>_{a_{k}}\otimes\left|1\right>_{b_{k}}\otimes\left|0\right>_{c_{k}}\equiv\left|E_{2}\right>_{k}\\
\left|E_{3}'\left(\tau\right)\right>_{k} & = & \left|0\right>_{a_{k}}\otimes\left[-\sin\left(\omega_{k}\tau\right)\left|0\right>_{b_{k}}\otimes\left|0\right>_{c_{k}}+\cos\left(\omega_{k}\tau\right)\left|1\right>_{b_{k}}\otimes\left|1\right>_{c_{i}}\right]\\
 &  & \equiv\left|0\right>_{a_{k}}\otimes\left|\text{H}_{\tau}^{\perp}\right>_{b_{k}c_{k}}\\
 &  & \equiv\cos\left(\omega_{k}\tau\right)\left|E_{3}\right>_{k}-\sin\left(\omega_{k}\tau\right)\left|E_{0}\right>_{k}\\
 &  & \quad=\left|E_{3}\right>_{k}-\omega_{k}\tau\left|E_{0}\right>_{k}+\mathcal{O}(\tau^{2})\\
\left|E_{4}'\left(\tau\right)\right>_{k} & = & \left|1\right>_{a_{k}}\otimes\left[\cos\left(\omega_{k}\tau\right)\left|0\right>_{b_{k}}\otimes\left|0\right>_{c_{k}}+\sin\left(\omega_{k}\tau\right)\left|1\right>_{b_{k}}\otimes\left|1\right>_{c_{k}}\right]\\
 &  & \equiv\left|1\right>_{a_{k}}\otimes\left|\text{H}_{\tau}\right>_{b_{k}c_{k}}\\
 &  & \equiv\cos\left(\omega_{k}\tau\right)\left|E_{4}\right>_{k}+\sin\left(\omega_{k}\tau\right)\left|E_{7}\right>_{k}\\
 &  & \quad=\left|E_{4}\right>_{k}+\omega_{k}\tau\left|E_{7}\right>_{k}+\mathcal{O}(\tau^{2})\\
\left|E_{5}'\right>_{k} & = & \left|0\right>_{a_{k}}\otimes\left|0\right>_{b_{k}}\otimes\left|1\right>_{c_{k}}\equiv\left|E_{5}\right>_{k}\\
\left|E_{6}'\right>_{k} & = & \left|0\right>_{a_{k}}\otimes\left|0\right>_{b_{k}}\otimes\left|1\right>_{c_{k}}\equiv\left|E_{6}\right>_{k}\\
\left|E_{7}'\left(\tau\right)\right>_{k} & = & \left|1\right>_{a_{k}}\otimes\left[-\sin\left(\omega_{k}\tau\right)\left|0\right>_{b_{k}}\otimes\left|0\right>_{c_{k}}+\cos\left(\omega_{k}\tau\right)\left|1\right>_{b_{k}}\otimes\left|1\right>_{c_{k}}\right]\\
 &  & \equiv\left|1\right>_{a_{k}}\otimes\left|\text{H}_{\tau}^{\perp}\right>_{b_{k}c_{k}}\\
 &  & \equiv\cos\left(\omega_{k}\tau\right)\left|E_{7}\right>_{k}-\sin\left(\omega_{k}\tau\right)\left|E_{4}\right>_{k}\\
 &  & \quad=\left|E_{7}\right>_{k}-\omega_{k}\tau\left|E_{4}\right>_{k}+\mathcal{O}(\tau^{2}),
\end{array}\label{eq:base_uprime}
\end{equation}
\foreignlanguage{american}{where $\left|\text{H}_{\tau}^{\perp}\right>$
denotes a Hawking state orthogonal to the Hawking state $\left|\text{H}_{\tau}\right>$.
Since, from a geometrical point of view, $\omega_{k}$ is an angle
of rotation in the 8-dimensional Hilbert space $\mathcal{H}_{k}$
for the $k$-mode sector, the auxiliary ``time'' parameter $\tau$
is dimensionless. More precisely, $\tau$ is a scale parameter governing
the evolution, and $\tau\in\left[0,1\right]$. In particular, for
$\tau=1$ the evolution is supposed to be completed.}

The third base, $\left\{ \left|E_{\Lambda}''\left(\tau\right)\right>_{k}\right\} _{\Lambda=0}^{7}$,
is given by: 
\begin{equation}
\begin{array}{rcl}
\left|E_{0}^{\prime\prime}\right>_{k} & = & \left|0\right>_{a_{k}}\otimes\left|0\right>_{b_{k}}\otimes\left|0\right>_{c_{k}}\equiv\left|E_{0}\right>_{k}\\
\left|E_{1}^{\prime\prime}\left(\tau\right)\right>_{k} & = & {\displaystyle \cos\left(\frac{\pi\tau}{2}\right)\left|0\right>_{a_{k}}\otimes\left|0\right>_{b_{k}}\otimes\left|1\right>_{c_{k}}-\sin\left(\frac{\pi\tau}{2}\right)\left|1\right>_{a_{k}}\otimes\left|0\right>_{b_{k}}\otimes\left|0\right>_{c_{k}}}\\
 &  & {\displaystyle \equiv\cos\left(\frac{\pi\tau}{2}\right)\left|E_{1}\right>_{k}-\sin\left(\frac{\pi\tau}{2}\right)\left|E_{4}\right>_{k}}\\
 &  & {\displaystyle \quad=\left|E_{1}\right>_{k}-\frac{\pi\tau}{2}\left|E_{4}\right>_{k}+\mathcal{O}(\tau^{2})}\\
\left|E_{2}^{\prime\prime}\right>_{k} & = & \left|0\right>_{a_{k}}\otimes\left|1\right>_{b_{k}}\otimes\left|0\right>_{c_{k}}\equiv\left|E_{2}\right>_{k}\\
\left|E_{3}^{\prime\prime}\right>_{k} & = & \left|0\right>_{a_{k}}\otimes\left|1\right>_{b_{k}}\otimes\left|1\right>_{c_{k}}\equiv\left|E_{3}\right>_{k}\\
\left|E_{4}^{\prime\prime}\left(\tau\right)\right>_{k} & = & {\displaystyle \sin\left(\frac{\pi\tau}{2}\right)\left|0\right>_{a_{k}}\otimes\left|0\right>_{b_{k}}\otimes\left|1\right>_{c_{k}}+\cos\left(\frac{\pi\tau}{2}\right)\left|1\right>_{a_{k}}\otimes\left|0\right>_{b_{k}}\otimes\left|0\right>_{c_{k}}}\\
 &  & {\displaystyle \equiv\cos\left(\frac{\pi\tau}{2}\right)\left|E_{4}\right>_{k}+\sin\left(\frac{\pi\tau}{2}\right)\left|E_{1}\right>_{k}}\\
 &  & {\displaystyle \quad=\left|E_{4}\right>_{k}+\frac{\pi\tau}{2}\left|E_{1}\right>_{k}+\mathcal{O}(\tau^{2})}\\
\left|E_{5}^{\prime\prime}\right>_{k} & = & \left|1\right>_{a_{k}}\otimes\left|0\right>_{b_{k}}\otimes\left|1\right>_{c_{k}}\equiv\left|E_{5}\right>_{k}\\
\left|E_{6}^{\prime\prime}\right>_{k} & = & \left|1\right>_{a_{k}}\otimes\left|1\right>_{b_{k}}\otimes\left|0\right>_{c_{k}}\equiv\left|E_{6}\right>_{k}\\
\left|E_{7}^{\prime\prime}\right>_{k} & = & \left|1\right>_{a_{k}}\otimes\left|1\right>_{b_{k}}\otimes\left|1\right>_{c_{k}}\equiv\left|E_{7}\right>_{k}.
\end{array}\label{eq:base_ebis}
\end{equation}

Now we define ($\tau$-dependent) unitary operators according to the
recipe ({\ref{eq:general_u}}) as follows: 
\begin{equation}
U_{k}^{\prime}\left(\tau\right)=\sum_{\Lambda=0}^{7}\left|E_{\Lambda}'\left(\tau\right)\right>_{k}\left<E_{\Lambda}\right|_{k},\label{eq:uprime_defined}
\end{equation}
\begin{equation}
U_{k}^{\prime\prime}\left(\tau\right)=\sum_{\Lambda=0}^{7}\left|E_{\Lambda}^{\prime\prime}\left(\tau\right)\right>_{k}\left<E_{\Lambda}^{\prime}\left(\tau\right)\right|_{k},\label{eq:ubis_defined}
\end{equation}
and

\begin{equation}
\begin{array}{rcl}
U_{k}\left(\tau\right) & \equiv & U_{k}^{\prime\prime}\left(\tau\right)\cdot U_{k}^{\prime}\left(\tau\right)={\displaystyle \sum_{\Lambda,\Lambda'=0}^{7}\left|E_{\Lambda}^{\prime\prime}\left(\tau\right)\right>\left<E_{\Lambda}'\left(\tau\right)|E_{\Lambda'}'\left(\tau\right)\right>\left<E_{\Lambda'}\right|}\\
 & = & {\displaystyle \sum_{\Lambda=0}^{7}\left|E_{\Lambda}^{\prime\prime}\left(\tau\right)\right>\left<E_{\Lambda}\right|.}
\end{array}\label{eq:u_defined}
\end{equation}
We can easily confirm that for $\tau=1$ the (explicitly) unitary
operators ({\ref{eq:uprime_defined}}), ({\ref{eq:ubis_defined}})
and ({\ref{eq:u_defined}}) act according to the rules ({\ref{eq:imode_primeu}}),
({\ref{eq:imode_bisu}}) and ({\ref{eq:imode_u}}), respectively.
For example, for $U_{k}'\left(\equiv U_{k}'\left(1\right)\right)$
we confirm that 
\begin{equation}
\begin{array}{c}
U'_{k}\left(\left|q_{k}\right>_{a_{k}}\otimes\left|\text{O}\right>_{b_{k}c_{k}}\right)=U'_{k}\left\{ \left[\left(1-q_{k}\right)\left|0\right>_{a_{k}}+q_{k}\left|1\right>_{a_{k}}\right]\otimes\left|\text{O}\right>_{b_{k}c_{k}}\right\} \\
={\displaystyle \sum_{\Lambda=0}^{7}\left|E_{\Lambda}'\left(1\right)\right>_{k}\left<E_{\Lambda}\right|_{k}\left[\left(1-q_{k}\right)\left|E_{0}\right>_{k}+q_{k}\left|E_{4}\right>_{k}\right]}\\
=\left(1-q_{k}\right)\left|E_{0}'\left(1\right)\right>_{k}+q_{k}\left|E_{4}'\left(1\right)\right>_{k}=\left[\left(1-q_{k}\right)\left|0\right>_{a_{k}}+q_{k}\left|1\right>_{a_{k}}\right]\otimes\left|\text{H}_{1}\right>_{b_{k}c_{k}}\\
=\left|q_{k}\right>_{a_{k}}\otimes\left|\text{H}_{1}\right>_{b_{k}c_{k}},
\end{array}\label{eq:example_uprime}
\end{equation}
as expected (see ({\ref{eq:imode_primeu}})).

\section{Physical picture}

Now, let us determine a corresponding (infinitesimal) generator $\mathbb{H}_{k}$
(``Hamiltonian'') for the unitary evolution operator $U_{k}\left(\tau\right)$,
i.e. 
\begin{equation}
U_{k}\left(\tau\right)=\mathbb{I}_{k}-i\tau\mathbb{H}_{k}+\mathcal{O}(\tau^{2}).\label{eq:definition_hk}
\end{equation}
To this end we will utilize expansions (in terms of the ``time''
parameter $\tau$) of the bases $\left\{ \left|E_{\Lambda}'\left(\tau\right)\right>\right\} $
and $\left\{ \left|E_{\Lambda}''\left(\tau\right)\right>\right\} $
given in ({\ref{eq:base_uprime}}) and ({\ref{eq:base_ebis}}),
respectively, around the standard base $\left\{ \left|E_{\Lambda}\right>\right\} $.

To make a comparison to a known case (i.e. to the fermion squeezing
operator \citep{svozil_1990,Khanna_Thermal}), let us derive the form
of the generator $\mathbb{H}_{k}'$ for the unitary transformation
$U_{k}'(\tau)$ ({\ref{eq:uprime_defined}}). By virtue of ({\ref{eq:uprime_defined}}),
({\ref{eq:base_uprime}}) and ({\ref{eq:definition_hk}}) (with
unprimed quantities replaced by primed ones) we get 
\begin{equation}
\begin{array}{rcl}
U_{k}'\left(\tau\right) & = & \mathbb{I}_{k}-i\tau i\omega_{k}\left(\left|E_{3}\right>_{k}\left<E_{0}\right|_{k}-\left|E_{0}\right>_{k}\left<E_{3}\right|_{k}\right.\\
 & + & \left.\left|E_{7}\right>_{k}\left<E_{4}\right|_{k}-\left|E_{4}\right>_{k}\left<E_{7}\right|_{k}\right)+\mathcal{O}\left(\tau^{2}\right).
\end{array}\label{eq:hprimek}
\end{equation}
Introducing the identification: 
\begin{equation}
\begin{array}{rcl}
\left|0\right>_{x_{k}}\left<0\right|_{x_{k}} & = & \hat{x}_{k}\hat{x}_{k}^{\dagger}\\
\left|0\right>_{x_{k}}\left<1\right|_{x_{k}} & = & \hat{x}_{k}\\
\left|1\right>_{x_{k}}\left<0\right|_{x_{k}} & = & \hat{x}_{k}^{\dagger}\\
\left|1\right>_{x_{k}}\left<1\right|_{x_{k}} & = & \hat{x}_{k}^{\dagger}\hat{x}_{k},
\end{array}\label{eq:fock_operators}
\end{equation}
for the modes $x_{k}=a_{k},b_{k},c_{k}$ ($k=1,2,\dots,N$), from
({\ref{eq:hprimek}}) we obtain a representation of the operator
$\mathbb{H}'_{k}$ in the Fock space, i.e. 
\begin{equation}
\hat{\mathbb{H}}_{k}'=i\omega_{k}\hat{b}_{k}^{\dagger}\hat{c}_{k}^{\dagger}+\text{H.c.},\label{eq:f_ham_prim}
\end{equation}
where ``H.c.'' means Hermitian conjugate. The operator ({\ref{eq:f_ham_prim}})
is known as a two-mode fermion squeezing operator, responsible for
creation of fermionic pairs in the framework of the Hawking effect
(see e.g.\ Sect. 5.2 in \citep{Mann:2015luq}).

Let us now derive the generator $\mathbb{H}_{k}$, and its Fock space
counterpart $\hat{\mathbb{H}}_{k}$, for the entire evolution $U_{k}(\tau)$.
By virtue of ({\ref{eq:u_defined}}), ({\ref{eq:base_ebis}})
and ({\ref{eq:definition_hk}}) we obtain 
\begin{equation}
U_{k}\left(\tau\right)=\mathbb{I}_{k}-i\tau\frac{i\pi}{2}\left(\left|E_{1}\right>_{k}\left<E_{4}\right|_{k}-\left|E_{4}\right>_{k}\left<E_{1}\right|_{k}\right)+\mathcal{O}\left(\tau^{2}\right).\label{eq:fock_h_total}
\end{equation}
Implementing the identification ({\ref{eq:fock_operators}}) we
obtain the corresponding Fock space generator (``Hamiltonian'')
\begin{equation}
\hat{\mathbb{H}}_{k}=\frac{i\pi}{2}\hat{a}_{k}\hat{b}_{k}\hat{b}_{k}^{\dagger}\hat{c}_{k}^{\dagger}+\text{H.c.}\label{eq:f_ham}
\end{equation}
One should note that the operator $\hat{\mathbb{H}}_{k}$ is 4-linear,
which should be contrasted with a bilinear structure of the squeezing
operator ({\ref{eq:f_ham_prim}}) and a trilinear structure of the
operator discussed in the context of the Hawking effect in \citep{Nation:2010va}.
According to ({\ref{eq:operatorsu_as_products}}) the total ``Hamiltonian''
is $\hat{\mathbb{H}}=\sum_{k=1}^{N}\hat{\mathbb{H}}_{k}$.

\begin{figure}
\centering{}\includegraphics[scale=0.17]{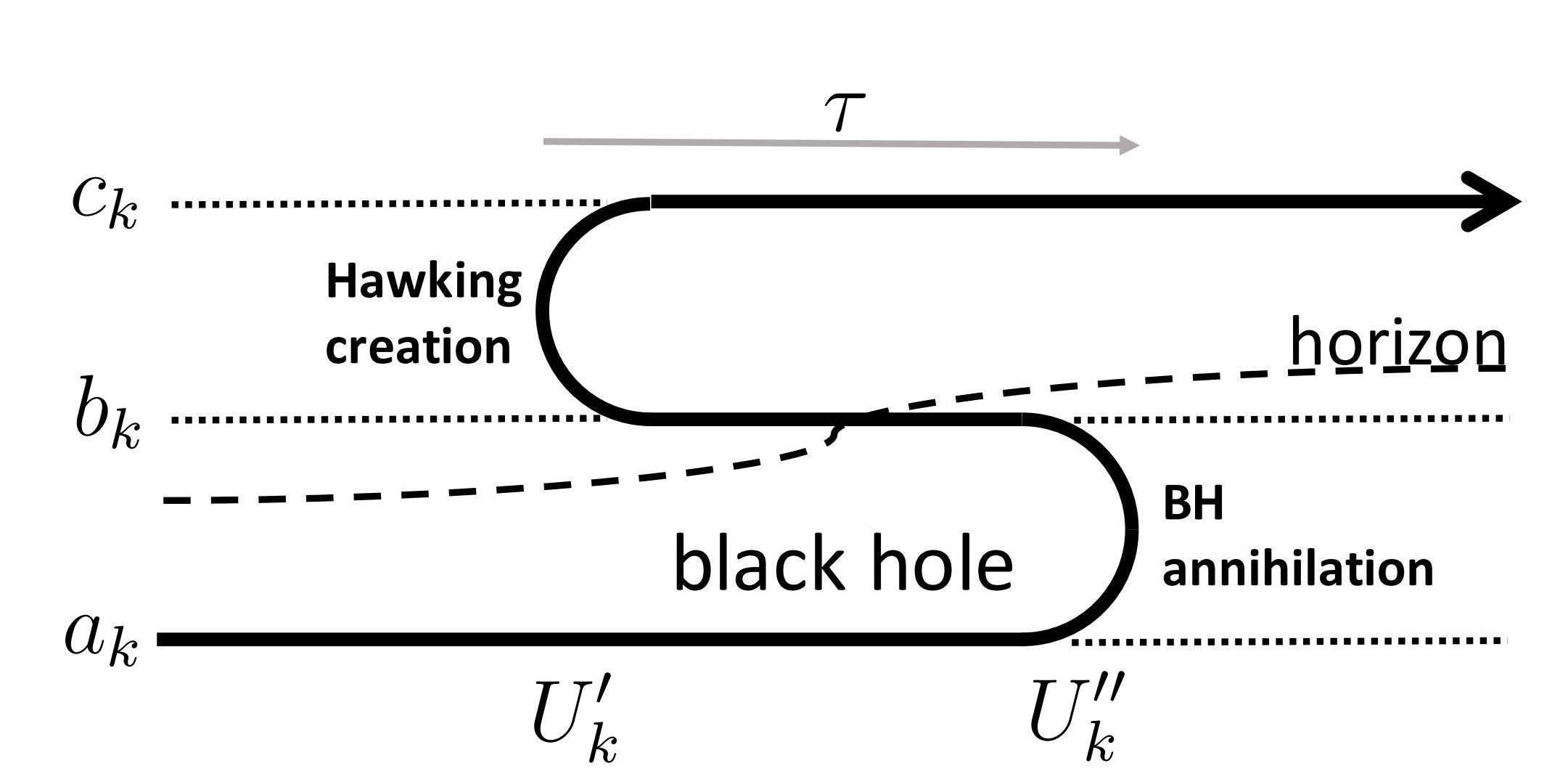} \caption{\label{fig:1}A Feynman-like diagram/line (the S-shaped thick line
with an arrow) depicts the entire qubit transport process $U_{k}$,
i.e. a composition of the Hawking particle pair creation $U'_{k}$
outside the BH and of particle pair annihilation $U''_{k}$ inside
the BH, in the spirit of a tunneling phenomenon.}
\end{figure}

Coming back to the global description of the BH unitary evolution,
we would like to draw the reader's attention to a possible interpretation,
depicted in Fig.~{\ref{fig:1}}. Namely, Fig.~{\ref{fig:1}},
in an intuitive way, presents the entire unitary process $U_{k}$
of qubit transfer from a BH to the outgoing radiation as a composition
of the two processes, $U'_{k}$ and $U''_{k}$, i.e. the Hawking particle
pair creation outside the BH and later particle pair annihilation
inside the BH, respectively. A Feynman-like diagram/line depicts the
transfer of a qubit in the spirit of a tunneling phenomenon.

\selectlanguage{american}%
Taking into account ``reversibility'' of unitary processes, the
Reader could rightfully expect that at some moment of the evolution
the ``reverse'' process of a given one should also occur, and a
``reverse diagram'' to the one presented in Fig.~1 should also
appear. The unitary transformations introduced in Section III are
actually rotations in 8-dimensional complex vector spaces with rotation
angles proportional to the ``time'' parameter $\tau$. Then, if
we start from a vector corresponding to an $a_{k}$-mode and ``rotate''
it onto a $c_{k}$-mode, and next we stop the evolution, we effectively
swap one particle for another. If we instead further continued the
evolution we would return to the $a_{k}$-mode back. Therefore, we
have to assume that the proposed evolution is only valid within a
limited period of ``time'' $\tau$, namely $\tau\in\left[0,1\right]$.
That limitation could be justified by the assumption that the evolution
is ``effective'' rather than ``fundamental'' in the sense that
its temporary form is determined by the current structure of spacetime
and the distribution of matter. Then, when $a_{k}$-modes become transformed
into $c_{k}$-modes, the situation changes and the evolution in the
proposed form switches off. Moreover, one should also note that the
whole result is influenced be the initial state of the system, i.e.\ we
start with modes only inside the black hole, while all other modes
are in the vacuum state. Such an assumption corresponds to the idealized
(model) situation assuming that in the beginning of the evolution
we only deal with a (single) black hole (represented by $a_{k}$-modes
occupied) and empty spacetime outside the black hole ($c_{k}$-modes
empty).
\selectlanguage{english}%

\section{Final remarks}

Primarily inspired by a recent paper of Osuga and Page \citep{Osuga2018},
in particular by their Eq.~(3.3), essentially the same as Eq.~(3.3)
in \citep{Almheiri2013} (accidental coincidence of the numbers of
the equations!), we have proposed a unitary toy model of BH evaporation,
which is an extension of the model introduced in \citep{Broda2020}.
By virtue of the construction the model is explicitly unitary and
it describes transport of qubits from a BH to the outgoing radiation.
For interpretational simplicity and a more direct relation to particle
language, we have decided to formulate our qubit model in terms of
fermions. As a byproduct of our construction, for possible reference
to other models of the Hawking effect and BH evaporation, besides
the global version of the evolution operator, we have determined its
infinitesimal form (``Hamiltonian''). In turn the global evolution,
involving the Hawking creation as well as latter annihilation inside
the BH, can intuitively be interpreted as a tunneling phenomenon as
depicted in Fig.~\ref{fig:1}.

 \bibliographystyle{plain}
\bibliography{fermionic_model}

\end{document}